\documentclass[12pt]{article}
\usepackage{amsmath}
\usepackage{cite}
\usepackage{geometry}
\usepackage{textcomp} 
\usepackage[colorlinks=true, linkcolor=blue, citecolor=blue, urlcolor=blue]{hyperref}
\title{\textbf{Critical Reflections on ``Overcoming a Challenge for Bohmian Mechanics'' by H. Nikoli\'c and the Experimental Findings of Sharoglazova \textit{et al.}}}
\author{
\large Mikołaj Sienicki\thanks{Polish-Japanese Academy of Information Technology, ul. Koszykowa 86, 02-008 Warsaw, Poland, European Union.} 
\quad and \quad
Krzysztof Sienicki\thanks{Chair of Theoretical Physics of Naturally Intelligent Systems (\textcopyright~NIS),
Lipowa 2/Topolowa 19, 05-807 Podkowa Leśna, Poland, European Union.}
}
\date{\today}

\begin{document}

\maketitle
\begin{abstract}
This paper offers a brief reflection on H. Nikolić’s response to the experimental findings of Sharoglazova et al., which challenge Bohmian mechanics. While Nikolić’s revision satisfies the continuity equation, it reintroduces assumptions he seeks to avoid and overlooks key empirical and nonlocal aspects of the system. These issues underscore unresolved tensions in applying Bohmian mechanics to complex, interacting regimes.
\end{abstract}

In a recent experimental study, Sharoglazova \textit{et al.}~\cite{sharoglazova2025} observed an unexpected energy--speed relationship in a quantum system composed of two coupled waveguides. Their results revealed a discrepancy between measured particle speeds and those predicted by standard Bohmian mechanics, which typically defines the velocity of a quantum particle as
\begin{equation}
v(x,t) = \frac{\hbar}{m} \nabla \varphi(x,t),
\end{equation}
where \( \varphi(x,t) \) is the phase of the wave function \( \psi = \sqrt{\rho}\, e^{i\varphi} \). This formula, foundational to Bohmian theory as articulated by Bohm~\cite{bohm1952} and Holland~\cite{holland1993}, assumes that the system evolves under a standard single-component Schrödinger equation.

In response, Nikoli\'c~\cite{nikolic2025matters} argues that the challenge posed by~\cite{sharoglazova2025} does not undermine Bohmian mechanics. He observes that the coupled-mode system studied in the experiment does not satisfy the standard Schrödinger equation structure required for Eq.~(1) to hold. Instead, Nikoli\'c proposes a generalization based on the continuity equation,
\begin{equation}
\frac{\partial \rho}{\partial t} + \nabla \cdot (\rho \mathbf{v}) = 0,
\end{equation}
which governs the conservation of probability. In the one-dimensional, time-independent case relevant to the experiment, this reduces to
\begin{equation}
\frac{d}{dx} \left[ \rho(x) v_x(x) \right] = 0 \quad \Rightarrow \quad v_x(x) = \frac{c}{\rho(x)}.
\end{equation}

Nikoli\'c then derives a velocity field of the form
\begin{equation}
v_x(x) = \frac{1}{L \rho(x)} \cdot \frac{\hbar k_2}{m},
\end{equation}
arguing that this resolves the inconsistency by properly modeling the transport in the waveguide system while satisfying the continuity condition.

Although Nikolić’s mathematical formulation is undoubtedly solid—Equation (4) does satisfy the continuity equation—his reasoning leaves several important questions hanging. Most notably, the velocity field he proposes depends heavily on a normalization condition that, somewhat paradoxically, invokes the average of the very expression (Equation 1) he seeks to dismiss in this context. This kind of circular logic weakens the claim that Equation (4) arises cleanly from first principles.

More significantly, the solution he puts forward feels more like a tactical fix than a comprehensive, predictive model grounded in the broader architecture of Bohmian mechanics—particularly when dealing with composite or multi-mode systems. It’s a patch, not a programmatic foundation.

There’s also a deeper empirical issue flagged by Sharoglazova \textit{et al.}~\cite{sharoglazova2025}, which his reply largely avoids. If experimentally measured values like energy and velocity repeatedly diverge from Bohmian predictions, the discrepancy isn’t just philosophical—it’s experimental. Nikolić argues that Bohmian velocities aren’t directly observable and thus any such mismatches don’t threaten the theory’s validity. But in a realist interpretation like Bohmian mechanics, even unobservable quantities must yield predictions that are indirectly testable and empirically reliable. Failing that, the theory risks losing its physical plausibility.

Another concern lies in his treatment of particle dynamics, which assumes that each waveguide can be considered in isolation using effective one-particle densities. This simplification downplays the potentially nonlocal or entangled structure of the full quantum state. In coupled systems, more holistic approaches—such as those based on configuration space or conditional wave functions—might offer a stronger theoretical footing~\cite{durr1992,nikolic2005,durr2009}.

To sum up, while Nikolić~\cite{nikolic2025matters} provides a careful and mathematically consistent rejoinder to the experimental results reported by Sharoglazova \textit{et al.}~\cite{sharoglazova2025}, his response does not fully grapple with the broader implications raised by the data. The real issue is not whether Bohmian mechanics can be retrofitted to accommodate the system, but whether it can offer a coherent and predictive account of quantum transport in complex, interacting settings without sacrificing its interpretational integrity. As it stands, the critique by Sharoglazova \textit{et al.}~\cite{sharoglazova2025} remains both compelling and unresolved, warranting continued scrutiny of Bohmian mechanics’ empirical and theoretical boundaries.

\end{document}